\documentclass[12pt,english]{article}
\pdfoutput=1
\usepackage{graphicx,array}
\usepackage{cite}
\usepackage{color}
\usepackage{latexsym}
\usepackage{amsthm}
\usepackage{amsmath}
\usepackage[titletoc]{appendix}
\usepackage{enumitem}
\usepackage{amssymb}
\usepackage[unicode=true,
 bookmarks=true,bookmarksnumbered=false,bookmarksopen=false,
 breaklinks=false,pdfborder={0 0 1},backref=false,colorlinks=true]
 {hyperref}
\hypersetup{pdftitle={The Hilbert Space of Quantum Gravity Is Locally Finite-Dimensional},
 pdfauthor={Ning Bao, Sean M. Carroll, and Ashmeet Singh},
 citecolor=blue,linkcolor=blue,urlcolor=blue,citecolor=blue,linkcolor=blue}
\usepackage{breakurl}
\usepackage[hang,flushmargin]{footmisc} 

\def\simgt{\mathrel{\lower2.5pt\vbox{\lineskip=0pt\baselineskip=0pt
           \hbox{$>$}\hbox{$\sim$}}}}
\def\simlt{\mathrel{\lower2.5pt\vbox{\lineskip=0pt\baselineskip=0pt
           \hbox{$<$}\hbox{$\sim$}}}}

\newcommand{\be}{\begin{equation}}
\newcommand{\ee}{\end{equation}}
\newcommand{\bea}{\begin{eqnarray}}
\newcommand{\eea}{\end{eqnarray}}

\newcommand{\Tr}{\operatorname{Tr}}

\newcommand{\QG}{\mathrm{QG}}
\newcommand{\EFT}{\mathrm{EFT}}

\newcommand{\sys}{\mathrm{sys}}
\newcommand{\env}{\mathrm{env}}
\newcommand{\HH}{\mathcal{H}}

\definecolor{nicered}{rgb}{0.7,0.1,0.1}
\definecolor{nicegreen}{rgb}{0.1,0.5,0.1}
\hypersetup{colorlinks,citecolor=black,linkcolor=black,urlcolor=black}

\begin{document}
\baselineskip=14pt
\hfill CALT-TH-2017-17
\hfill

\vspace{2cm}
\thispagestyle{empty}
\begin{center}
{\LARGE\bf
The Hilbert Space of Quantum Gravity\\
Is Locally Finite-Dimensional
}\\
\bigskip\vspace{1cm}{
{\large Ning Bao, Sean M.\ Carroll, and Ashmeet Singh}\footnote{e-mail: \url{ningbao75@gmail.com, seancarroll@gmail.com, ashmeet@theory.caltech.edu}}
} \\[7mm]
 {\it Walter Burke Institute for Theoretical Physics\\
    California Institute of Technology,
   Pasadena, CA 91125} \\
 \end{center}
\bigskip
\centerline{\large\bf Abstract}

\begin{quote} \small
We argue in a model-independent way that the Hilbert space of quantum gravity is locally finite-dimensional.
In other words, the density operator describing the state corresponding to a small region of space, when such a notion makes sense, is defined on a finite-dimensional factor of a larger Hilbert space.
Because quantum gravity potentially describes superpositions of different geometries, it is crucial that we associate Hilbert-space factors with spatial regions only on individual decohered branches of the universal wave function.
We discuss some implications of this claim, including the fact that quantum field theory cannot be a fundamental description of Nature.
\end{quote}

\vspace{3cm}

\begin{quote} \footnotesize
Essay written for the Gravity Research Foundation 2017 Awards for Essays on Gravitation.
\end{quote}

\newpage
\baselineskip=16pt
	
\setcounter{footnote}{0}

In quantum field theory, the von~Neumann entropy of a compact region of space $R$ is infinite, because an infinite number of degrees of freedom in the region are entangled with an infinite number outside.
In a theory with gravity, however, if we try to excite these degrees of freedom, many states collapse to black holes with finite entropy $S=A/4G$, where $A$ is the horizon area \cite{bekenstein1973, bekenstein1981}.
It is conceivable that there are degrees of freedom within a black hole that do not contribute to the entropy.
However, if such states were low-energy, the entropy of the black hole could increase via entanglement, violating the Bekenstein bound.
If they are sufficiently high-energy that they don't become entangled, exciting them would increase the size of the black hole, taking it out of the ``local region'' with which it was associated.

The finiteness of black hole entropy therefore upper bounds the number of degrees of freedom that can be excited within $R$, and therefore on the dimensionality of $\HH_R$, the factor of Hilbert space associated with $R$ (as the dimensionality of Hilbert space is roughly the exponential of the number of degrees of freedom).
Similarly, a patch of de~Sitter space, which arguably represents an equilibrium configuration of spacetime, has a finite entropy proportional to its horizon area, indicative of a finite-dimensional Hilbert space \cite{Fischler2000,Banks2000,bousso2000,Banks:2000fe,Witten:2001kn,Dyson2002,Parikh:2004wh,Carroll:2017kjo}.

The most straightforward interpretation of this situation is that in the true theory of nature, which includes gravity, any local region is characterized by a finite-dimensional factor of Hilbert space.
Some take this statement as well-established, while others find it obviously wrong.
Here we argue that the straightforward interpretation is most likely correct, even if the current state of the art prevents us from drawing  definitive conclusions.

This discussion begs an important question: what is ``the Hilbert space associated with a region"?
Quantum theories describe states in Hilbert space, and notions like ``space" and ``locality'' should emerge from that fundamental level \cite{vanRaamsdonk2010,Cao:2016mst,Cotler:2017abq}.
Our burden is therefore to understand what might be meant by the Hilbert space of a local region, and whether that notion is well-defined in quantum gravity.

We imagine that the fundamental quantum theory of nature describes a density operator $\rho$ acting on a Hilbert space $\HH_\QG$.
The entanglement structure of near-vacuum states in spacetime is very specific, so generic states in $\HH_\QG$ won't look like spacetime at all \cite{Cao:2016mst,Cotler:2017abq}.
Rather, in phenomenologically relevant, far-from-equilibrium states $\rho$, there will be macroscopic pointer states representing semiclassical geometries.
To that end, we imagine a decomposition
\be
  \HH_\QG = \HH_\sys \otimes \HH_{\env},
\ee
where the system factor $\HH_\sys$ will describe a region of space and its associated long-wavelength fields, and the environment factor $\HH_\env$ is traced over to obtain our system density matrix, $\rho_\sys = \Tr_\env \rho$.
The environment might include microscopic degrees of freedom that are either irrelevant or spatially distant.
Then decoherence approximately diagonalizes the system density matrix in the pointer basis,
\be
  |\Psi_a\rangle \in \HH_\sys, \quad \rho_\sys = \sum_a p_a |\Psi_a\rangle\langle\Psi_a|.
\ee

\begin{figure}[t]
\centering
\includegraphics[width=.95\textwidth]{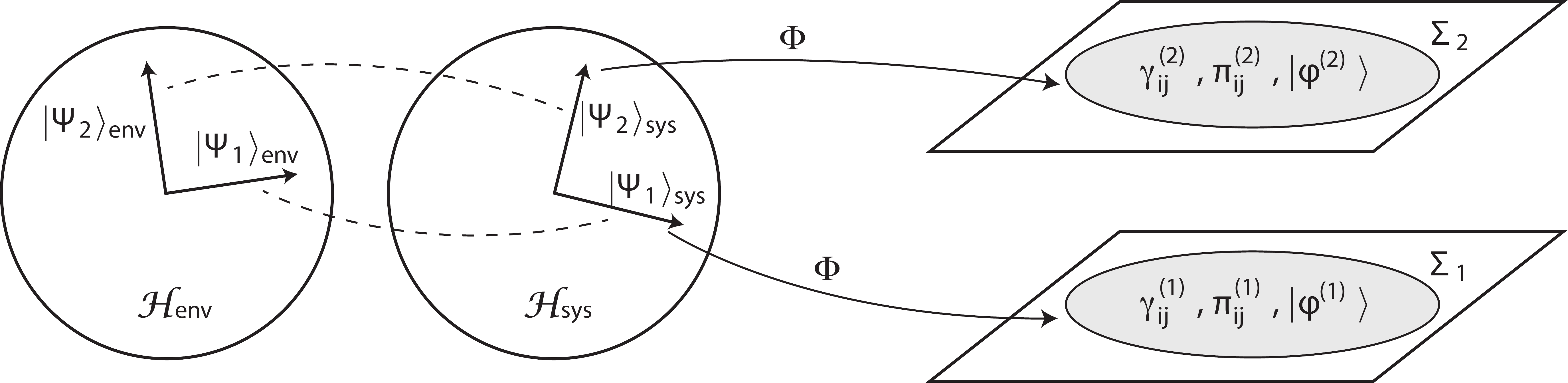}
\caption{On the left, the Hilbert space of quantum gravity is portrayed as a tensor product, $\HH_\QG = \HH_\sys \otimes \HH_{\env}$, and two decohered branches $|\Psi_a\rangle_{\mathrm{sys}}\otimes |\Psi_a\rangle_{\mathrm{env}}$ are shown. Under the map $\Phi$, the system factors of these states map to regions $\Sigma_a$ of a semiclassical spacetime background, on which are defined a spatial metric and its conjugate momentum as well as the quantum state of an effective field theory.}
\label{fig:qse}
\end{figure}

An emergence map $\Phi$ associates decohered branches of the quantum-gravity wave function with states of an effective field theory defined on a semiclassical background spacetime.
In a space+time decomposition, the emergent theory describes spatial manifolds $\Sigma$ with 3-metric $\gamma_{ij}$ and conjugate momentum $\pi_{ij}$, along with a Hilbert space $\HH_\EFT^{(\Sigma)}$ of quantum fields on $\Sigma$,
\be
  \Phi\, :\, |\Psi_a\rangle \rightarrow \left\{\Sigma_a, \gamma_{ij}^{(a)}, \pi_{ij}^{(a)}, |\phi^{(a)}\rangle\right\},
\ee
where $|\phi^{(a)}\rangle \in \HH_\EFT^{(\Sigma_a)}$.
We don't insist that $\Sigma_a$ be a boundaryless surface; it may simply be a finite region of space (e.g., the interior of a de~Sitter horizon).
Horizon complementarity \cite{Susskind:1993if} suggests that there could be a limit on the extent of $\Sigma_a$, perhaps in the form of an entropy bound \`a la Bousso \cite{bousso1999}.

Now imagine that, in some emergent geometry, we divide $\Sigma$ into a closed region $R$ and its exterior $\bar{R}$.
The proposition ``a region of space is described by a finite-dimensional factor of Hilbert space" should be interpreted as the claim that we can decompose the system Hilbert space in the fundamental theory as $\HH_\sys = \HH_R \otimes \HH_{\bar R}$, representing factors describing physics inside and outside $R$, such that $\dim \HH_R$ is finite.
(As is standard with emergent theories, the map $\HH_R \otimes \HH_{\bar R} \rightarrow \HH_R^{(\Sigma)} \otimes \HH_{\bar R}^{(\Sigma)}$ represents an approximation. In particular, our ability to divide space into two sets of degrees of freedom doesn't imply that we can continue to subdivide it into many small regions simultaneously.)

Such a decomposition is familiar in the case of quantum field theories on a fixed spacetime background.
(There may be subtleties due to gauge invariance on the boundary, e.g. \cite{Casini:2013rba,Harlow:2015lma}.)
This is necessitated by the success of locality as an underlying principle of everyday physics.
If a single degree of freedom were accessible both in $R$ and $\bar R$, a unitary operator localized within $R$ could change the state elsewhere -- not in the sense of branching the wave function, but in the sense of direct superluminal information transfer, as the reduced density matrix $\rho_{\bar R}=\Tr_R\rho_\sys$ would change instantly.
Indeed, the notion that we can sensibly talk about the entropy of a black hole implicitly assumes this kind of locality.

A complication could arise due to the fact that gravity is a long-range force coupled to a conserved charge (mass-energy) that is always positive \cite{marolf2015,giddings2015, donnelly+giddings2016_1, donnelly+giddings2016_2}.
Any operator with nonzero energy or other conserved Poincar\'e charges is ``dressed'' by a gravitational field stretching out to infinity, representing an apparent obstacle to localization. 
In particular, diffeomorphism-invariant operators are considered from the start.

A locally-finite theory, however, need not give rise \emph{directly} to a diffeomorphism-invariant description of gravity.
The fundamental description could correspond to a particular gauge, in which the symmetries of the theory weren't manifest, even though they could be restored once the effective theory had emerged.
It is dangerous to start with symmetries of the sought-after continuum theory, defined in the context of an infinite-dimensional Hilbert space, and demand that they be present at the discrete level; all that we should require is that crucial physical properties ultimately emerge (e.g. \cite{Barcelo:2016xhp}).

From our perspective, gravitational dressing is not inconsistent with localizing a finite number of degrees of freedom within a region of space.
(Even the notion of ``at infinity'' is unlikely to be well-defined as a statement about $\HH_\QG$, but we won't rely on that loophole.)
Rather, there are superselection sectors corresponding to total gauge charges, including energy; operators that change the total energy connect one quantum-gravity pointer state to a state with a different energy.
Such operators are not necessary for describing the local dynamics, which can be captured entirely by operators that commute with global charges; physical changes in the local state of a system are not instantaneously communicated to infinity.
Equivalently, there is no obstacle to decomposing the Hamiltonian into a term acting only within $R$, one acting only within $\bar R$, and an interaction defined by a sum of tensor products of such local operators.
When working in any one such sector, then, we are allowed to factorize our system Hilbert space as $\HH_\sys = \HH_R \otimes \HH_{\bar R}$.

Given the validity of this factorization, the finite dimensionality of $\HH_R$ follows from the black-hole arguments above, with one  consideration: when we attempt to excite degrees of freedom until we reach a black hole, we shouldn't consider operators that change the Poincar\'e charges at infinity.
This is no problem; it is easy enough to imagine creating black holes simply by moving around existing mass/energy within the state (as actually happens in the real world when black holes are created).
The upper bound from black-hole entropy on the number of ways this could happen implies an upper bound on the effective degrees of freedom, and therefore on $\dim \HH_\sys$.
 
We therefore conclude that it is sensible to associate factors of Hilbert space with regions of space, at the level of individual branches representing semiclassical spacetimes, and that such factors have finite dimensionality.
There are well-known obstacles to constructing phenomenological acceptable theories with finite-dimensional Hilbert spaces, including a lack of exact Lorentz invariance.
It is therefore imperative to investigate whether such features can arise in an approximate fashion without violating experimental bounds \cite{mattingly2005}.

Locally finite-dimensional Hilbert spaces entail a number of consequences.
This perspective suggests new tools for investigating the behavior of quantum spacetime, such as the quantum-circuit approach \cite{bao_etal2017}.
It also implies an attitude toward ultraviolet divergences in quantum gravity: there are no such divergences, as there are only finitely many degrees of freedom locally.
The cosmological constant problem becomes the question of why the factor associated with our de~Sitter patch has its particular large, finite dimensionality, $\exp(10^{122})$ \cite{Banks:2000fe}; perhaps other fine-tuning questions, such as the hierarchy problem, can be similarly recast.

\begin{center} 
 {\bf Acknowledgments}
 \end{center}

We thank Scott Aaronson, Steven Giddings, and John Preskill for helpful discussions. This research is funded in part by the Walter Burke Institute for Theoretical Physics at Caltech, by DOE grant DE-SC0011632, and by the Foundational Questions Institute.

\bibliographystyle{utphys}
\bibliography{locallyfiniterefs}

\providecommand{\href}[2]{#2}\begingroup\raggedright\begin{thebibliography}{10}

\bibitem{bekenstein1973}
J.~D. Bekenstein, ``Black holes and entropy,''
  \href{http://dx.doi.org/10.1103/PhysRevD.7.2333}{{\em Phys. Rev. D}
  {\bfseries 7} (Apr, 1973) 2333--2346}.
  \url{http://link.aps.org/doi/10.1103/PhysRevD.7.2333}.

\bibitem{bekenstein1981}
J.~D. Bekenstein, ``Universal upper bound on the entropy-to-energy ratio for
  bounded systems,'' \href{http://dx.doi.org/10.1103/PhysRevD.23.287}{{\em
  Phys. Rev. D} {\bfseries 23} (Jan, 1981) 287--298}.
  \url{http://link.aps.org/doi/10.1103/PhysRevD.23.287}.

\bibitem{Fischler2000}
W.~Fischler, ``{Taking de Sitter Seriously}.'' Talk given at Role of Scaling
  Laws in Physics and Biology (Celebrating the 60th Birthday of Geoffrey West),
  Santa Fe, Dec., 2000.

\bibitem{Banks2000}
T.~Banks, ``{QuantuMechanics and CosMology}.'' Talk given at the festschrift
  for L. Susskind, Stanford University, May 2000, 2000.

\bibitem{bousso2000}
R.~Bousso, ``Positive vacuum energy and the n-bound,'' {\em Journal of High
  Energy Physics} {\bfseries 2000} no.~11, (2000) 038,
  \href{http://arxiv.org/abs/https://arxiv.org/abs/hep-th/0010252}{{\ttfamily
  https://arxiv.org/abs/hep-th/0010252}}.
  \url{http://stacks.iop.org/1126-6708/2000/i=11/a=038}.

\bibitem{Banks:2000fe}
T.~Banks, ``{Cosmological breaking of supersymmetry?},''
  \href{http://dx.doi.org/10.1142/S0217751X01003998}{{\em Int. J. Mod. Phys.}
  {\bfseries A16} (2001) 910--921},
\href{http://arxiv.org/abs/hep-th/0007146}{{\ttfamily arXiv:hep-th/0007146
  [hep-th]}}.

\bibitem{Witten:2001kn}
E.~Witten, ``{Quantum gravity in de Sitter space},'' in {\em {Strings 2001:
  International Conference Mumbai, India, January 5-10, 2001}}.
\newblock 2001.
\newblock \href{http://arxiv.org/abs/hep-th/0106109}{{\ttfamily
  arXiv:hep-th/0106109 [hep-th]}}.
\newblock
\url{http://alice.cern.ch/format/showfull?sysnb=2259607}.
\newblock

\bibitem{Dyson2002}
L.~Dyson, M.~Kleban, and L.~Susskind, ``{Disturbing implications of a
  cosmological constant},''
  \href{http://dx.doi.org/10.1088/1126-6708/2002/10/011}{{\em JHEP} {\bfseries
  10} (2002) 011}, \href{http://arxiv.org/abs/hep-th/0208013}{{\ttfamily
  arXiv:hep-th/0208013}}.

\bibitem{Parikh:2004wh}
M.~K. Parikh and E.~P. Verlinde, ``{De Sitter holography with a finite number
  of states},'' \href{http://dx.doi.org/10.1088/1126-6708/2005/01/054}{{\em
  JHEP} {\bfseries 01} (2005) 054},
\href{http://arxiv.org/abs/hep-th/0410227}{{\ttfamily arXiv:hep-th/0410227
  [hep-th]}}.

\bibitem{Carroll:2017kjo}
S.~M. Carroll and A.~Chatwin-Davies, ``{Cosmic Equilibration: A Holographic
  No-Hair Theorem from the Generalized Second Law},''
\href{http://arxiv.org/abs/1703.09241}{{\ttfamily arXiv:1703.09241 [hep-th]}}.

\bibitem{vanRaamsdonk2010}
M.~Van~Raamsdonk, ``{Building up spacetime with quantum entanglement},''
  \href{http://dx.doi.org/10.1007/s10714-010-1034-0,
  10.1142/S0218271810018529}{{\em Gen. Rel. Grav.} {\bfseries 42} (2010)
  2323--2329}, \href{http://arxiv.org/abs/1005.3035}{{\ttfamily arXiv:1005.3035
  [hep-th]}}.
[Int. J. Mod. Phys.D19,2429(2010)].

\bibitem{Cao:2016mst}
C.~Cao, S.~M. Carroll, and S.~Michalakis, ``{Space from Hilbert Space:
  Recovering Geometry from Bulk Entanglement},''
  \href{http://dx.doi.org/10.1103/PhysRevD.95.024031}{{\em Phys. Rev.}
  {\bfseries D95} no.~2, (2017) 024031},
\href{http://arxiv.org/abs/1606.08444}{{\ttfamily arXiv:1606.08444 [hep-th]}}.

\bibitem{Cotler:2017abq}
J.~S. Cotler, G.~R. Penington, and D.~H. Ranard, ``{Locality from the
  Spectrum},''
\href{http://arxiv.org/abs/1702.06142}{{\ttfamily arXiv:1702.06142
  [quant-ph]}}.

\bibitem{Susskind:1993if}
L.~Susskind, L.~Thorlacius, and J.~Uglum, ``{The Stretched horizon and black
  hole complementarity},''
  \href{http://dx.doi.org/10.1103/PhysRevD.48.3743}{{\em Phys. Rev.} {\bfseries
  D48} (1993) 3743--3761},
\href{http://arxiv.org/abs/hep-th/9306069}{{\ttfamily arXiv:hep-th/9306069
  [hep-th]}}.

\bibitem{bousso1999}
R.~Bousso, ``{A Covariant entropy conjecture},''
  \href{http://dx.doi.org/10.1088/1126-6708/1999/07/004}{{\em JHEP} {\bfseries
  07} (1999) 004},
\href{http://arxiv.org/abs/hep-th/9905177}{{\ttfamily arXiv:hep-th/9905177
  [hep-th]}}.

\bibitem{Casini:2013rba}
H.~Casini, M.~Huerta, and J.~A. Rosabal, ``{Remarks on entanglement entropy for
  gauge fields},'' \href{http://dx.doi.org/10.1103/PhysRevD.89.085012}{{\em
  Phys. Rev.} {\bfseries D89} no.~8, (2014) 085012},
\href{http://arxiv.org/abs/1312.1183}{{\ttfamily arXiv:1312.1183 [hep-th]}}.

\bibitem{Harlow:2015lma}
D.~Harlow, ``{Wormholes, Emergent Gauge Fields, and the Weak Gravity
  Conjecture},'' \href{http://dx.doi.org/10.1007/JHEP01(2016)122}{{\em JHEP}
  {\bfseries 01} (2016) 122},
\href{http://arxiv.org/abs/1510.07911}{{\ttfamily arXiv:1510.07911 [hep-th]}}.

\bibitem{marolf2015}
D.~Marolf, ``{Emergent Gravity Requires Kinematic Nonlocality},''
  \href{http://dx.doi.org/10.1103/PhysRevLett.114.031104}{{\em Phys. Rev.
  Lett.} {\bfseries 114} no.~3, (2015) 031104},
\href{http://arxiv.org/abs/1409.2509}{{\ttfamily arXiv:1409.2509 [hep-th]}}.

\bibitem{giddings2015}
S.~B. Giddings, ``{Hilbert space structure in quantum gravity: an algebraic
  perspective},'' \href{http://dx.doi.org/10.1007/JHEP12(2015)099}{{\em JHEP}
  {\bfseries 12} (2015) 099},
\href{http://arxiv.org/abs/1503.08207}{{\ttfamily arXiv:1503.08207 [hep-th]}}.

\bibitem{donnelly+giddings2016_1}
W.~Donnelly and S.~B. Giddings, ``{Diffeomorphism-invariant observables and
  their nonlocal algebra},''
  \href{http://dx.doi.org/10.1103/PhysRevD.94.029903,
  10.1103/PhysRevD.93.024030}{{\em Phys. Rev.} {\bfseries D93} no.~2, (2016)
  024030}, \href{http://arxiv.org/abs/1507.07921}{{\ttfamily arXiv:1507.07921
  [hep-th]}}.
[Erratum: Phys. Rev.D94,no.2,029903(2016)].

\bibitem{donnelly+giddings2016_2}
W.~Donnelly and S.~B. Giddings, ``{Observables, gravitational dressing, and
  obstructions to locality and subsystems},''
  \href{http://dx.doi.org/10.1103/PhysRevD.94.104038}{{\em Phys. Rev.}
  {\bfseries D94} no.~10, (2016) 104038},
\href{http://arxiv.org/abs/1607.01025}{{\ttfamily arXiv:1607.01025 [hep-th]}}.

\bibitem{Barcelo:2016xhp}
C.~Barceló, R.~Carballo-Rubio, F.~Di~Filippo, and L.~J. Garay, ``{From
  physical symmetries to emergent gauge symmetries},''
  \href{http://dx.doi.org/10.1007/JHEP10(2016)084}{{\em JHEP} {\bfseries 10}
  (2016) 084},
\href{http://arxiv.org/abs/1608.07473}{{\ttfamily arXiv:1608.07473 [gr-qc]}}.

\bibitem{mattingly2005}
D.~Mattingly, ``{Modern tests of Lorentz invariance},''
  \href{http://dx.doi.org/10.12942/lrr-2005-5}{{\em Living Rev. Rel.}
  {\bfseries 8} (2005) 5},
\href{http://arxiv.org/abs/gr-qc/0502097}{{\ttfamily arXiv:gr-qc/0502097
  [gr-qc]}}.

\bibitem{bao_etal2017}
N.~Bao, C.~Cao, S.~M. Carroll, and L.~McAllister, ``{Quantum Circuit Cosmology:
  The Expansion of the Universe Since the First Qubit},''
\href{http://arxiv.org/abs/1702.06959}{{\ttfamily arXiv:1702.06959 [hep-th]}}.

\end{thebibliography}\endgroup

\end{document}